\documentclass[aps,prl,preprint,groupedaddress]{revtex4-2}
\usepackage{graphicx}
\usepackage{amsmath}
\usepackage{mathtools}
\usepackage{amssymb}
\DeclareMathOperator{\sech}{sech}
\begin{document}


\title{Electron Acceleration during Macroscale Magnetic Reconnection}


\author{H. Arnold}
\affiliation{IREAP, University of Maryland, College Park MD 20742-3511, USA}
\author{J.~F.~Drake}
\affiliation{IREAP, University of Maryland, College Park MD 20742-3511, USA}
\author{M.~Swisdak}
\affiliation{IREAP, University of Maryland, College Park MD 20742-3511, USA}
\author{F.~Guo}
\affiliation{Los Alamos National Lab, Los Alamos NM 87545, USA}
\author{J. T. Dahlin}
\affiliation{NASA Goddard Space Flight Center, Greenbelt MD 20771, USA}
\author{B. Chen}
\affiliation{New Jersey Institute of technology, Newark NJ 07102, USA}
\author{G. Fleishman}
\affiliation{New Jersey Institute of technology, Newark NJ 07102, USA}
\author{L. Glesener}
\affiliation{University of Minnesota, Minneapolis MN 55455, USA}
\author{E. Kontar}
\affiliation{University of Glasgow, Glasgow, UK, G12 8QQ}
\author{T. Phan}
\affiliation{University of California, Berkeley CA, USA}
\author{C. Shen}
\affiliation{Harvard University, Cambridge MA, USA}


\date{\today}

\begin{abstract}
The first self-consistent simulations of electron acceleration during
magnetic reconnection in a macroscale system are presented. Consistent
with solar flare observations the spectra of energetic electrons take
the form of power-laws that extend more than two decades in
energy. The drive mechanism for these nonthermal electrons is Fermi
reflection in growing and merging magnetic flux ropes. A strong guide
field is found to suppress the production of nonthermal electrons by
weakening the Fermi drive mechanism. For a weak guide field the total
energy content of nonthermal electrons dominates that of the hot
thermal electrons even though their number density remains small. Our
results are benchmarked with the hard x-ray, radio and extreme
ultra-violet (EUV) observations of the X8.2-class solar flare on
September 10, 2017.

\end{abstract}


\maketitle

{\bf Introduction} Solar flares are explosive events in the solar
corona that convert magnetic energy into particle
energy through magnetic reconnection
\cite{Lin1971,Emslie2004,Emslie2005,Emslie2012}. While some released
magnetic field energy goes into bulk flows and thermal energy, a
significant fraction appears in nonthermal electrons, which form a
power-law tail in the distribution function
\cite{Heristchi1992,Emslie2004,Emslie2012}. In some cases the pressure
from these electrons can approach the ambient magnetic pressure
\cite{Krucker2010,Oka2013}.  Observations by the Reuven Ramaty High
Energy Solar Spectroscopic Imager (RHESSI) and the Atmospheric Imaging
Assembly on the Solar Dynamic Observatory, suggest that these
energetic electrons make up a significant fraction of the total
electron density in the above-the-loop-top sources in solar flares
\cite{Krucker2014}.

Magnetic reconnection creates bent field lines with a tension that
drives an exhaust away from the x-line near the Alfv\'en speed
\cite{Parker1957,Lin1993}, energizing the surrounding plasma. When
particles stream into the exhaust, they gain energy by reflecting off
the bent field lines \cite{Drake2006b}. This process is referred
to as Fermi reflection and is believed to be responsible for producing
the power-law tails in the electron distribution function
\cite{Drake2006b,Drake2013,Guo2014,Li2019}. The energy gain of an
electron due to Fermi reflection is proportional to its energy
and therefore dominates the energy gain of the most energetic
electrons. Fermi reflection does not depend on any kinetic length
scale. Rather the relevant length is the curvature of the
reconnecting magnetic field. Electrons can continue to gain
energy from Fermi reflection even as flux ropes merge and approach the
domain size \cite{Fermo2010}.

This paper presents the first results of simulations of particle
acceleration during magnetic reconnection in a 2D macrosystem that
includes the self-consistent feedback of energetic particles on the
dynamics. The simulation model, {\it kglobal}
\cite{Drake2019,Arnold2019}, retains Fermi reflection as the dominant
drive mechanism for energetic electrons but excludes parallel electric
fields in kinetic-scale boundary layers.  Consequently, the
kinetic scales that constrain PIC modeling of macroscale systems are eliminated. For the first time we simulate magnetic
reconnection and particle acceleration in domains with sizes that
characterize energy release in solar flares. {\it kglobal} includes
the self-consistent feedback of energetic electrons on the large-scale
dynamics of reconnection so that the global energy is conserved.
While the ordering of {\it kglobal} eliminates kinetic-scale parallel
electric fields, the model includes large-scale parallel electric
fields, important for the initial energy gain of electrons during
reconnection \cite{Egedal2012,Egedal2015,Haggerty2015}.

The {\it kglobal} simulations produce power-law spectra of energetic
electrons that extend nearly three decades in energy and simultaneously produce super-hot thermal electrons that characterize flare
observations \cite{Lin2003,Krucker2010,Oka2013}. Consistent with
observations, the total energy content of the nonthermal electrons can
exceed that of the hot thermal electrons even though the number
density of the nonthermals is less than the hot thermals.
Simulations carried out with a variety of values of the initial
ambient out-of-plane guide field reveal that the strength of the guide
field strongly impacts the energy content of the nonthermal electrons
and their power-law index. A guide field exceeding the reconnecting
magnetic field suppresses production of nonthermal electrons by
weakening the Fermi drive mechanism. In contrast, the size of the
global system has relatively little influence on the production of
nonthermal electrons.
 
{\bf Numerical Simulations Setup} The 2D simulations presented
here are carried out with the {\it kglobal} model, which consists of a
magnetohydrodynamic (MHD) backbone with fluid ions and electrons as
well as particle electrons that are distributed as macroparticles on
the MHD grid. The two electron species combine so that charge
neutrality is preserved at all times
\cite{Drake2019,Arnold2019}. However, since the equations governing energy gain in the electron fluid are incomplete (e.g., Fermi reflection is not included), any change in the energy of the electron fluid will be neglected in the analysis of the electron energy gain.  The upstream reconnection magnetic field,
$B_0$, and the ion density, $n_0$, define the Alfv\'en speed,
$C_{A0}=B_0/\sqrt{4\pi m_in_0}$. Since no kinetic scales are resolved,
lengths are normalized to an arbitrary macroscale $L_0$. Times are
normalized to $\tau_A=L_0/C_{A0}$ and temperatures and particle energies
to $m_iC_{A0}^2$. The perpendicular electric field
follows an MHD scaling, $C_{A0}B_0/c$.  The parallel electric field
scales like $m_i C_{A0}^2/L_0e$ and is small compared with the
perpendicular component. However, the energy associated with the
parallel potential drop acting over the scale $L_0$ is of order
$m_iC_{A0}^2$, which is comparable to the available magnetic energy
per particle.

The simulations are initialized with constant densities and
pressures in a force-free current sheet and
periodic boundary conditions. Thus, $\boldsymbol{B}=B_0
\tanh{(y/w)}\boldsymbol{\hat{x}}+\\ \sqrt{B_0^2 \sech^2{(y/w)}+B_g^2} \boldsymbol{\hat{z}}$. The temperatures of all three
species are uniform and isotropic with $T_i=T_{e,part}=T_{e,fluid}=0.0625
m_iC_{A0}^2$, leading to an initial plasma $\beta$ of $0.25$
(based on $B_0$). While the initial $\beta$ is higher than typical coronal values, electron heating and acceleration is insensitive to this choice as well as to the chosen fraction of particle electrons ($25\%$ of the total electrons). The
domain size for all simulations is $L_x \times L_y = 2\pi L_0 \times
\pi/2 L_0$. The magnetic field evolution equation includes a
hyper-resistivity $\nu$ to facilitate reconnection, while minimizing
dissipation at large scales \cite{Kaw1979}. The effective Lundquist
number $S_\nu=C_AL_0^3/\nu$ associated with this hyper-resistivity is
varied to change the effective system size (ratio of the macro to
the dissipation scale). We also include fourth and second order
viscosity terms and some electron particle diffusion to prevent a
numerical instability associated with trapping electrons in small
perpendicular electric field fluctuations. Reconnection begins from
particle noise and proceeds to produce multiple flux ropes whose
number depends on $S_\nu$, with larger values of $S_\nu$ producing
more initial flux ropes. However, our late-time results are relatively
insensitive to $S_\nu$ and therefore the effective system size. Thus,
unless otherwise stated we focus on simulations with 100 particles per cell, time step $dt=0.0001 \tau_0$,
$S_\nu=9.5\times 10^7$, and $N_x\times N_y=2048 \times 512$ grid
cells. By taking $L_0=10^4km$, our grid cell is 30 km across, much larger than any kinetic scales or PIC simulations. The mass ratio is
$m_i/m_e=25$. The results are not sensitive to this value.  The speed
of light is $c/C_{A0} \approx 60$. We use guide fields $B_g/B_0=0.1$,
$0.25$, $0.4$, $0.5$, $0.6$, $0.8$, and $1.0$.

\begin{figure}
\centering
\includegraphics[width=32pc,height=22pc]{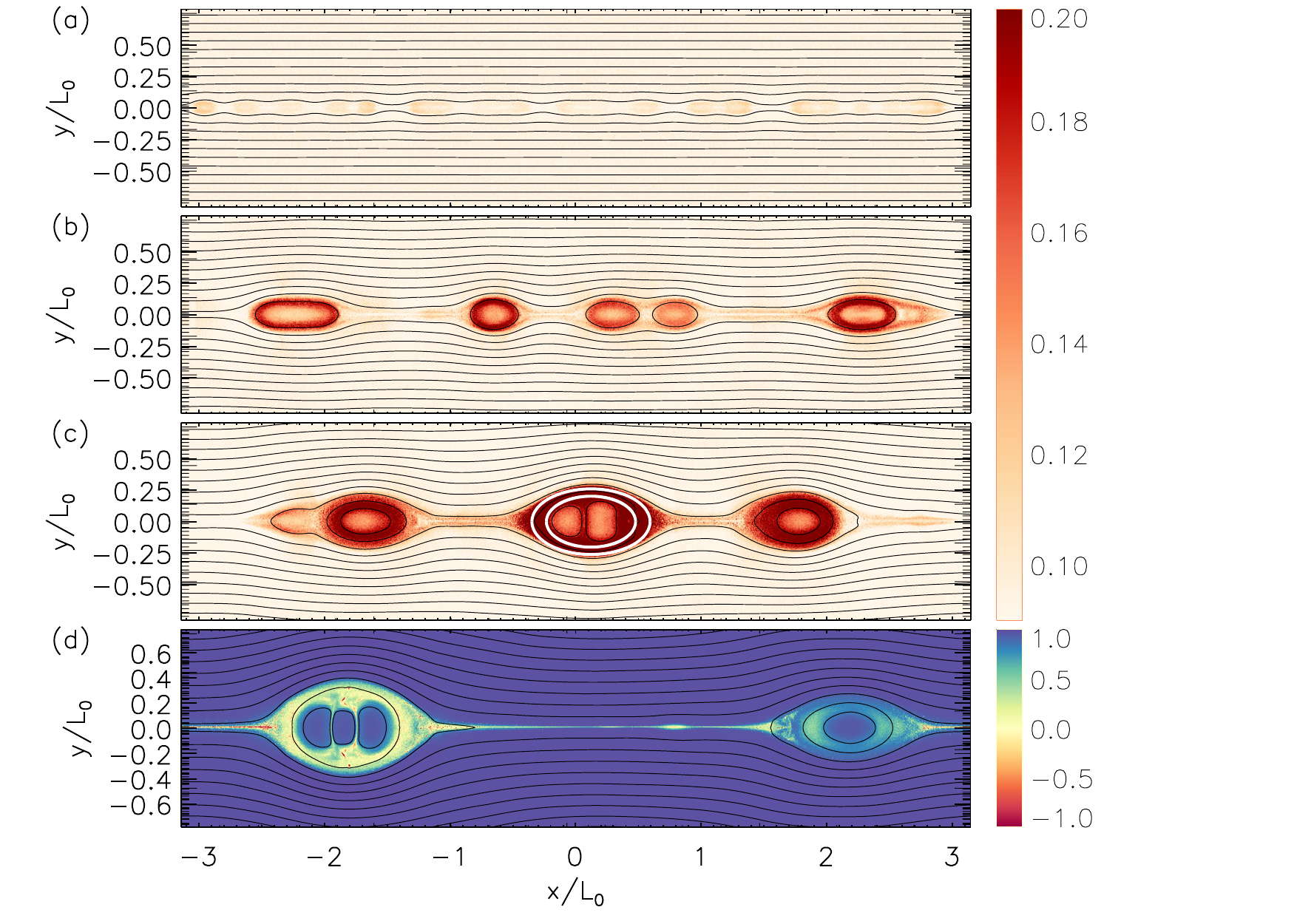}
\caption{In (a), (b), and (c): $<W>$ for a simulation with $B_g/B_0=0.25$
   in the $x-y$ plane at
  $t/\tau_A= 2.5$, $5$, and $8$ with magnetic field lines
  overplotted. In (d): the firehose stability parameter at late time
  from a simulation with $B_g/B_0=0.1$.}
\label{Ttot}
\end{figure}

{\bf Simulation Results} Since magnetic reconnection in our
simulations is triggered by particle noise, the dynamics begins with
the growth of many small islands, which subsequently undergo mergers
and eventually approach the system scale. This behavior is seen in
Fig.~\ref{Ttot} from a simulation with $B_g/B_0=0.25$. The energy per
particle of the particle electrons, $<W>$ (energy density divided by
number density), is shown in the $x-y$ plane at three times, $t/\tau_A
= 2.5$, $5$, and $8$ in panels (a), (b), and (c). Magnetic field lines
are superimposed. The particle electron energy is nearly constant
along field lines because of the high electron mobility parallel to
the magnetic field.

The firehose parameter, $1-4\pi(P_{||}-P_\perp)/ B^2$, is plotted in
Fig.~\ref{Ttot}(d) at late time from a simulation with
$B_g/B_0=0.1$. Large regions within the magnetic islands
are near marginal stability (with some unstable regions) so that the local
magnetic tension, which drives particle energy gain, is largely
suppressed within flux ropes in this simulation. Thus,
electron feedback on the MHD fluid is essential in regimes where
electron energy gain is significant. Models based on test particle
dynamics neglect the feedback of particles on the dynamics and can
therefore lead to runaway electron energy gain.

\begin{figure}
\centering
\includegraphics[width=32pc,height=22pc]{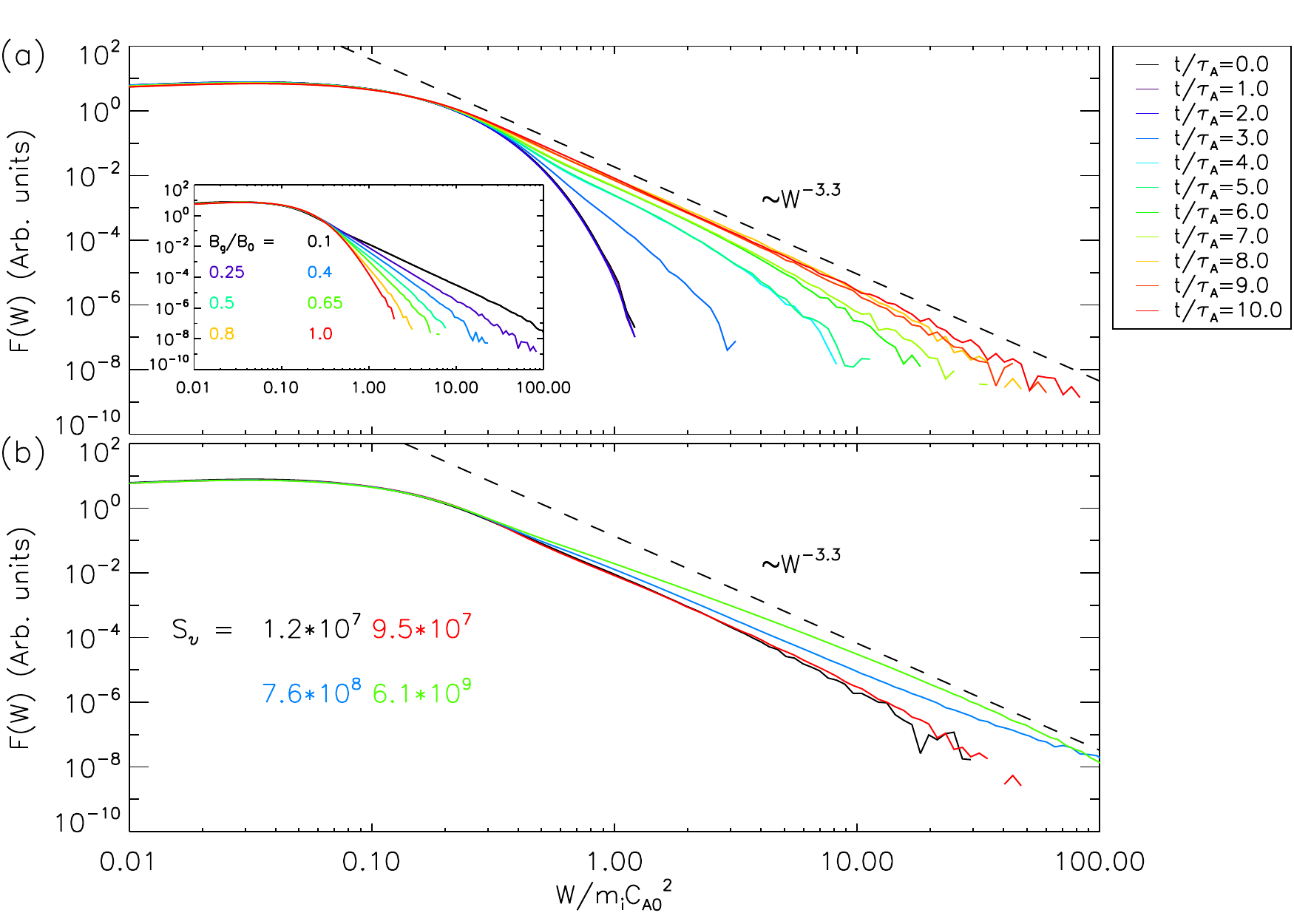}
\caption{In (a): a log-log plot of 
  $F(W)$ versus energy at multiple times for the $B_g/B_0=0.25$
  simulation. Inset in (a): the late time $F(W)$ for several guide
  fields. In (b): the late time $F(W)$ for $B_g/B_0=0.25$ with
  various values of $S_\nu$ (effective system size). The dashed line
  in both (a) and (b) is a power-law with $\delta'=3.3$.}
\label{distsPL}
\end{figure}

The spectrum of nonthermal electrons is calculated from the
simulations by summing the total number of particle electrons within a
specified energy range over the entire simulation domain. This ensures
that we maximize the count rate of the electrons at the highest
energies to improve the statistics of the measured distribution.  In
Fig.~\ref{distsPL}(a) we plot the differential electron number density
$F(W)=dN(W)/dW$ versus the normalized energy, $W/m_iC_{A0}^2$, on a
log-log scale at several times for the case $B_g/B_0=0.25$ shown in
Fig.~\ref{Ttot}.  $F(W)$ takes the form of a power-law (a straight
line in the log-log plot) as time progresses. The power-law index
$\delta '$ reaches a constant value at low energy early and extends to
higher energy over time.

The inset in Fig.~\ref{distsPL} shows the late-time $F(W)$ for several
values of the guide field corresponding to times during the
simulations when approximately the same amount of magnetic flux has
reconnected. As the guide field decreases, $\delta '$ decreases so the
spectrum becomes harder and more high-energy electrons are
produced. In Fig.~\ref{distsPL}(b) we plot the late time spectrum of
$F(W)$ for several values of $S_\nu$. Larger values of $S_\nu$
correspond to larger systems. Thus, Fig.~\ref{distsPL}(b) demonstrates
that the slope of the power-law of nonthermal electrons is relatively
insensitive to the size of the system. However, the total
energy contained in the nonthermal electrons increases with
reconnected flux, and thus a larger system produces a more extended
power-law.

\begin{figure}
\centering
\includegraphics[width=32pc,height=22pc]{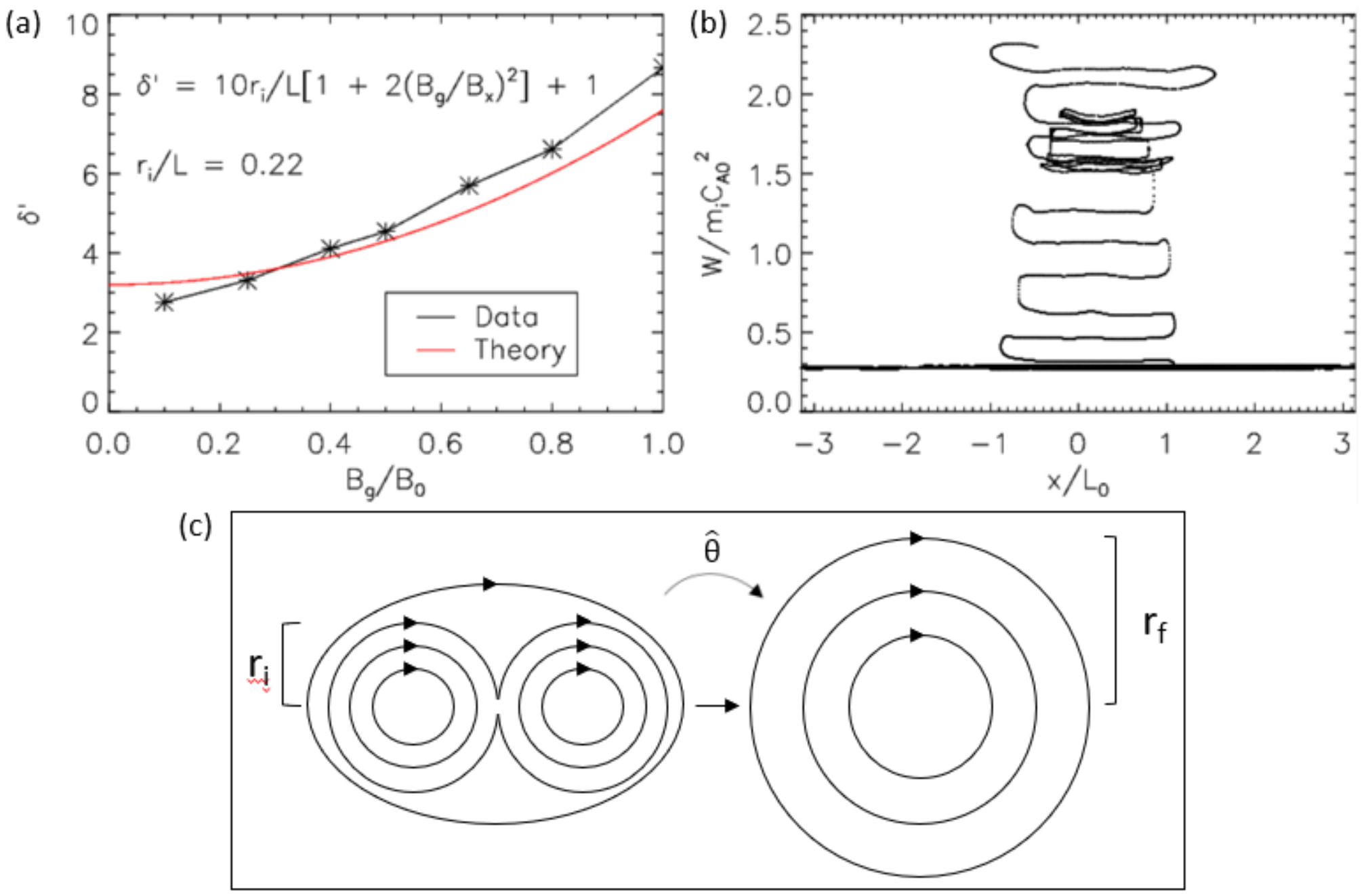}
\caption{In (a): $\delta'$ (black) versus guide field and the fit from the
  theoretical model (red). In (b): the energy versus $x$
  position of an electron that becomes a part of the nonthermal
  distribution. In (c): a schematic depicting the flux rope merging
  mechanism that leads to electron power-law distributions.}
\label{EvsX}
\end{figure}

The dependence of $\delta'$ on the guide field is
plotted in the curve marked by the stars in Fig.~\ref{EvsX}(a). A strong
guide field produces a soft nonthermal particle spectrum. The solid
red curve is from the theoretical model discussed in the next section.
Plotted in Fig.~\ref{EvsX}(b) is the time dependence of the energy of
a typical electron that populates the power-law tail versus the
$x$-position from a simulation with $B_g/B_0=0.25$. Early in time the
electron makes several passes through the system with little change in
energy. Once reconnection produces flux ropes, the electron is
captured by a flux rope and, as it contracts and merges with other
flux ropes, the electron undergoes Fermi reflection, gaining energy
with each bounce. Fig.~\ref{EvsX}(c) is a schematic of the
island-merging process that leads to the power-law tail (discussed in
the following section).

In exploring the power-law distribution of nonthermal electrons, we
averaged over the entire computational domain to improve the
statistics of the number of electrons with very high energy. However,
an important question in understanding particle energy gain during
reconnection concerns the relative numbers and energy content of
nonthermal electrons (those in the power-law tail) versus those that
display a thermal or nearly thermal distribution. The observations
suggest that the nonthermals often contain more energy than the hot
thermals in large flares\cite{Emslie2012, Aschwanden2016,
  Warmuth2016}. To explore these questions we analyzed data from more
limited spatial domains that include both hot thermal and nonthermal
electrons but exclude electrons that have not yet gained energy from
reconnection. We focus, therefore, on the interior of magnetic islands
where the electron temperature has increased and where there are
significant numbers of nonthermal electrons. The goals are to
establish whether a characteristic effective temperature is associated
with the hot thermal electrons and what fraction of the electrons can
be categorized as hot thermal versus nonthermal.

Specifically, we explore the region between the two white ellipses
within the middle flux rope in Fig. \ref{Ttot}(c). In
Fig.~\ref{en&den}(a) we show $F(W)$ (black line)
from the region between the two ellipses. The high-energy electrons
form a power-law distribution even in this localized region within a
single flux rope. In Fig.~\ref{en&den}(b) we display the same data, but on
a linear-linear scale focused on the lower energies to reveal the hot
thermal population. These two plots reveal that localized regions
within magnetic islands contain a mixture of electrons with
a range of energies so that the characteristics of the hot thermal
and nonthermals can be explored.

To model the distributions in Fig.~\ref{en&den} we use the sum of a
Maxwellian and a kappa distribution. The kappa
distribution fits the power-law tail of nonthermal electrons and the 
Maxwellian supplements the Maxwellian component of the
kappa distribution at low energy, producing a good fit to the hot
thermal electrons. The fitting procedure is discussed in detail
in the Supplementary section \cite{Supplement}. The outputs of the fit to the
electron distribution is the spectral index of the nonthermal
electrons, and the number density and total energy content of the nonthermal
and thermal electrons. 

The results of fitting for all of the guide fields appear in
Fig.~\ref{en&den}, (c) and (d). Shown in (c) is the percentage of the
total density (red) and energy (black) of the nonthermal electrons as
defined in \cite{Supplement} as a function of the guide
field. Each distribution that formed the basis of this data came from
a region within an island similar to the one shown in
Fig.~\ref{Ttot}(c). For a small guide field the energy content of the
nonthermal electrons is $\sim 80\%$ of the total particle electron
energy and $\sim 20\%$ of the total electron particle density. As the
guide field increases, the number of nonthermal electrons and their
energy content becomes small. In (d) is the total energy per particle
of the particle electrons (black) and the corresponding energy per
particle, or $(3/2)T_{th}$, of the hot thermal electrons (red), with
the initial energy shown as a dotted line. This is further evidence
that the nonthermal electrons dominate the total electron energy at
low guide fields where the Fermi drive is strong. On the other hand,
the energy of the hot thermal electrons is relatively insensitive to
the guide field and is likely controlled by slow shocks
(see \cite{Supplement}) rather than by Fermi reflection.

\begin{figure}
\centering
\includegraphics[width=32pc,height=22pc]{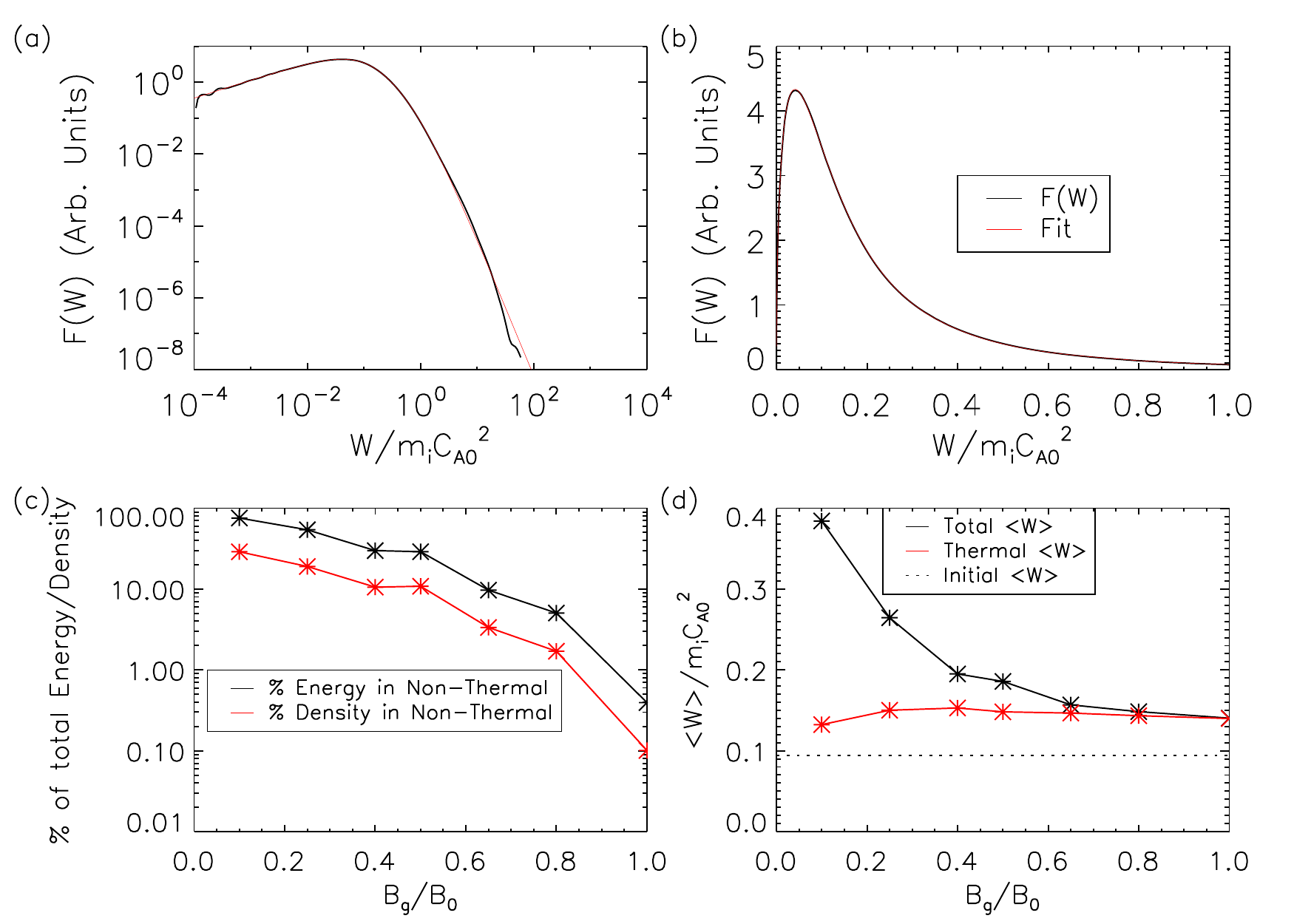}
\caption{In (a): $F(W)$ (black) along with the
  fit (red) described in the Supplementary section \cite{Supplement} versus energy
  on a log-log scale. In (b): the same data on a linear-linear
  scale, zoomed in to low energies to reveal the hot thermal
  electrons. In (c): the percentage of energy (black) and density
  (red) of nonthermal electrons versus guide
  field. In (d): the average energy per particle of particle
  electrons (black) and thermal electrons (red) versus
  guide field. The dotted line is the energy from the
  initial Maxwellian distribution of particle electrons.}
\label{en&den}
\end{figure}

{\bf An analytic model for nonthermal electron acceleration} We
present a model for electron acceleration in a current layer with
merging magnetic flux ropes that captures the essential results of the
{\it kglobal} simulations, including an expression for the power-law
index of the nonthermal electrons and its dependence on the ambient
guide field. The model includes the convective loss of electrons
injected into large, inactive flux ropes.

The model is based on electron energy gain during the merging of flux
ropes. The dominant heating, parallel to the local magnetic field
\cite{Dahlin2014,Dahlin2017}, results from the shortening of field
lines during flux rope mergers \cite{Drake2013} as shown in
Fig.~\ref{EvsX}(c): merging field lines contract from the figure-eight
configuration on the left to the circle on the right. Parallel heating
results from the invariance of the parallel action $\oint v_\parallel
dl$. Thus, the change in the energy during the merger of two flux
ropes can be calculated by evaluating the geometry of the magnetic
field before and after the merger. The calculation presented in the
Supplementary section \cite{Supplement}, results in the rate of energy gain
\begin{equation}
  \overset{\boldsymbol{.}}{W}=\frac{d}{dt}W=W\frac{g}{\tau_r}, 
  \label{energydot}
\end{equation}
with $\tau_r\sim r_i/Rc_{Ax}$ the merger time of a flux rope of
initial radius $r_i$, where $R\sim 0.1$ is the normalized rate of
merger of the flux ropes in the current layer, and $c_{Ax}$ is the Alfv\'en
speed based on the reconnecting magnetic field $B_x$. The factor $g=(1+2B_g^2/B_x^2)^{-1}$ arises from the dependence of
the radius of curvature of the reconnecting magnetic field on the
guide field \cite{Drake2006b,Dahlin2016,Dahlin2017}.
With the energy gain in Eq.~(\ref{energydot}),  an
equation can be derived for the number density $F(x, W, t)$ of electrons
per unit energy undergoing reconnection-driven acceleration in a
one-dimensional current layer and experiencing convective loss,
\begin{equation}
  \frac{\partial}{\partial t}F+\frac{\partial}{\partial x}v_x(x)F+\frac{\partial }{\partial W}\overset{\boldsymbol{.}}{W}F-D\frac{\partial^2}{\partial x^2}F=\frac{1}{\tau_{up}}F_{up}
  \label{fdot}
  \end{equation}
where $v_x(x)$ describes the convective loss of electrons as they are
ejected at the Alfv\'en speed out of the current layer and we include
a simple constant diffusion of electrons within the current layer. The
electrons are injected into the layer with an initial distribution
$F_{up}$ which is taken as a low-temperature Maxwellian. Although the
simulations carried out here are periodic and particles are therefore
not lost, the large flux ropes that emerge at late time and no longer
participate in the reconnection process act as sinks for energetic
electrons \cite{Dahlin2015, Li2019}. For a low upstream temperature and strong diffusion $D$, the steady state solution to Eq.~(\ref{fdot}) is given by
\begin{equation}
  F_0\propto \frac{1}{W}W^{-c_{Ax}\tau_r/gL} \sim W^{-(1+r_i/gRL)}.
  \label{f0scaling}
\end{equation}
The energetic electron spectrum is a power-law with a spectral index
that depends on the rate of reconnection $R$, the ratio of the
characteristic radius $r_i$ to the half-width $L$ of the current layer
and the relative strength of the guide field. A strong guide field for
which $g\sim (2B_g^2/B_x^2)^{-1}$ produces a very soft electron
spectrum. This scaling relation is compared with data from our
simulations in Fig.~\ref{EvsX}(a). The best fit of the model with the data has
$r_i/L=0.22$, which is consistent with the typical scale of islands in the simulations.

{\bf Comparison with Observations} The standard model for a solar
flare comes from \cite{Lin2000}. It includes an erupting flux rope
that produces a large reconnecting current sheet with a cusp-shaped
flare arcade below. As reconnection proceeds, more small flux ropes
are produced in the current sheet and flow either up toward the
erupting flux rope, or down toward the arcade. The solar flare of
September 10, 2017, was observed by several instruments
\cite{Gary2018,Warren2018,Chen2020,Fleishman2020}. The gyrosynchrotron
spectrum revealed relativistic electrons throughout the reconnecting
current sheet, with an increase in intensity where the current sheet
meets the arcade. The observed power-law indices, $\delta `$, for this
region fell in the range $3.5$-$6.5$, depending on the position. The RHESSI
observations for this event revealed both a footpoint and an extended
coronal source \cite{Gary2018}. The coronal source had a photon
spectral index near $4.4$, which for thin target emission corresponds
to a particle spectral index $\delta '=3.9$. Late in the flare the
temperature of the hot thermal electrons in the coronal current sheet
was analyzed with the EIS Fe XXIV/Fe XXIII ratio
\cite{Warren2018}. The hot thermal electron temperature had a broad
peak near 2.5keV.

Our simulations reveal that the spectral index of nonthermal electrons
and the number of the nonthermals depends strongly on the ambient
guide field. Comparisons of MHD simulations of the September 10, 2017,
flare with gyrosynchrotron emission of energetic electrons revealed
that the guide field was $30\%$ of the reconnecting magnetic field
\citep{Chen2020}. Based on the data from Fig.~\ref{distsPL}, our
simulations predict a power-law index near $3.5$, within the range
measured from gyrosynchrotron emission and very close to the RHESSI
measurement.

We can also compare the temperature jump of the hot thermal electrons
in our simulation with the measured 2.5keV from the EIS data. The
simulation data in Fig.~\ref{en&den}(d) suggests that the hot thermal
electrons should have a temperature jump near $0.04m_iC_{A0}^2$. The
Alfv\'en speed was estimated as 6,000 to 10,000 $km/s$
\citep{Chen2020}, which yields a temperature of 15-42$keV$, well above
the observations. On the other hand, such high flow speeds were not
measured. At the time of the EIS observations the measured outflow speed
in the current sheet was $\sim 800$km/s \cite{Cheng2018}, which is
likely a lower bound because of projection effects. Further, {\it in
  situ} measurements at 1AU indicate that the outflow speed is around
2/3 of the upstream Alfv\'en speed \cite{Haggerty2018}. Using
$1200$km/s for $C_{A0}$, we calculate a hot electron temperature of
$0.6keV$, which is smaller than the EIS measurement. Thus, more
precise measurements of the outflow speed are needed.

Observations of large numbers of flares have revealed that the energy
in nonthermal electrons exceeds that of the thermal electrons in $\sim 80\%$ of
events, suggesting
that solar flares are extremely efficient at accelerating nonthermal
electrons \cite{Aschwanden2016}. The efficiency of nonthermal electron
acceleration was greatest in large flares \cite{Warmuth2016} although
recent evidence from NuSTAR suggests that such results
might extend to smaller flares \citep{Glesener2020}.  Such results are
consistent with Fig.~\ref{en&den}(c) if $B_g/B_0<0.4$. The
simulations (Fig.~\ref{distsPL}(b)) further suggest that even small
flares might be efficient sources of nonthermal electrons.

An important question is why conventional PIC modeling has been unable
to produce the extended power-law spectra that develop in kglobal. Some
groups have been able to produce power-laws, such as \citep{Baumann2013} and \citep{Li2019}, however, they only extend about a decade in energy and they do not establish the dependence on the guide field. We
suggest that the short power-laws seen in PIC is because of inadequate separation between kinetic
scales and the macro-system scale. As electrons gain energy from Fermi
reflection in the PIC model, their Larmor radii become comparable to
the reconnecting flux ropes in the system and they become unmagnetized
\citep{Li2019}. This suppresses energy gain associated with Fermi
reflection \citep{Drake2013}. In PIC simulations the problem is extreme
because the macroscale in flares is of the order of $10^4km$ while the
simulation domains are of the order of 100's of meters. The
artificially large electron masses in PIC models further exacerbate
the problem. In a real flaring system the Larmor radii of electrons
with energies of 10's of MeV remain small so electron demagnetization
in reconnecting flux ropes is not likely to be important. In kglobal
the reconnecting flux ropes are at the macroscale and the electron
dynamics is guiding center and therefore fully magnetized.


\begin{acknowledgments}
The collaboration leading to these results was
facilitated by the NASA Drive Science Center on Solar Flare Energy
Release (SolFER), grant 80NSSC20K0627. We would like to thank
Dr.~W.~Daughton and participants in the NASA Drive Center SolFER for
invaluable discussions that contributed to this work.  This work has
been supported by NSF Grant Nos. PHY1805829 and PHY1500460 and the
FIELDS team of the Parker Solar Probe (NASA contract NNN06AA01C) and
the FINESST grant 80NSSC19K1435.  Fan Guo acknowledges support in part
from NASA grant 80NSSC20K1318 and Astrophysics Theory Program, and DOE
support through the LDRD program at LANL. Joel Dahlin was supported by
an appointment to the NASA Postdoctoral Program at the NASA Goddard
Space Flight Center, administered by Universities Space Research
Association under contract with NASA. The simulations were carried out
at the National Energy Research Scientific Computing Center
(NERSC). The data used to perform the analysis and construct the
figures for this paper are preserved at the NERSC High Performance
Storage System and are available upon request.
\end{acknowledgments}

\section{Supplemental}
{\bf Electron energy gain} As has been reported in previous PIC simulations we have monitored the
three mechanisms by which the particle electrons gain energy as a
function of time: Fermi reflection, the parallel electric field, and
betatron acceleration \cite{Dahlin2014}.  The data for guide fields
$B_g/B_0=1.0$ and $0.1$ are shown in Fig.~\ref{heating}(a) and (b),
respectively.  As in earlier simulations, electron energy gain is
bursty, which reflects the periodic merger of finite size flux
ropes. In the case of a weak guide field Fermi reflection dominates
energy gain during the entire simulation while acceleration by the
betatron mechanism (corresponding to the conservation of the magnetic
moment $V_\perp^2/B$) and the parallel electric field are
negligible. For a strong guide field betatron acceleration is again
negligible while acceleration by the parallel electric field becomes
comparable to the Fermi mechanism at late time. Notably, there is an
increase in heating due to $E_{||}$ toward the end of the simulation
for $B_g/B_0=1$. This is likely due to the development of the large
scale $E_{||}$ that forms after the electrons injected into the
reconnection exhaust gain significant energy. At this point the
resulting potential drop can heat electrons entering the exhaust as
documented in PIC simulations \cite{Egedal2012,Haggerty2015}. This
effect is subdominant in comparison to Fermi reflection for the case
of a small guide field. Note, however, that the overall electron
heating rate for the strong guide field case is more than an order of
magnitude lower than in the case of a weak guide field.

In Fig.~\ref{heating}(c) the parallel electric field is shown at late
time. The large-scale electric field points away from the current
sheet in the outflow exhausts as expected since the parallel electric
field serves to prevent hot electrons from escaping upstream
\cite{Egedal2012,Egedal2015,Haggerty2015}.  To see that the electric
field points away from the current sheet, note that $B_x$ is positive
above the current sheet and negative below so that $B_y$ is positive
on the left side of the left-most flux rope. Thus, $E_\parallel\sim
E_y$ is positive above the current sheet and negative below. The
unusual vertically oriented structures in $E_\parallel$ correspond to
the locations of slow shocks that propagate to the left and right in
the simulation. The {\it kglobal} model correctly describes the
potential drop across the slow shock that maintains charge
neutrality. The slow shocks produced during reconnection are not
effective in driving the nonthermal electrons
\cite{Zhang2019}. However, the electron distribution across the slow shock
reveals electron reflection and acceleration that likely causes heating of 
the thermal electrons and leads to the thermal energy gain seen in Fig. 4(d). This heating can also be observed in the upstream region in Fig.~\ref{heating}(d), which shows the average parallel energy per particle electron and has been overexposed to show the upstream heating associated with the slow shocks. This heating process will be explored in a future paper.

\begin{figure}
\setcounter{figure}{4}
\centering
\includegraphics[width=24pc,height=22pc]{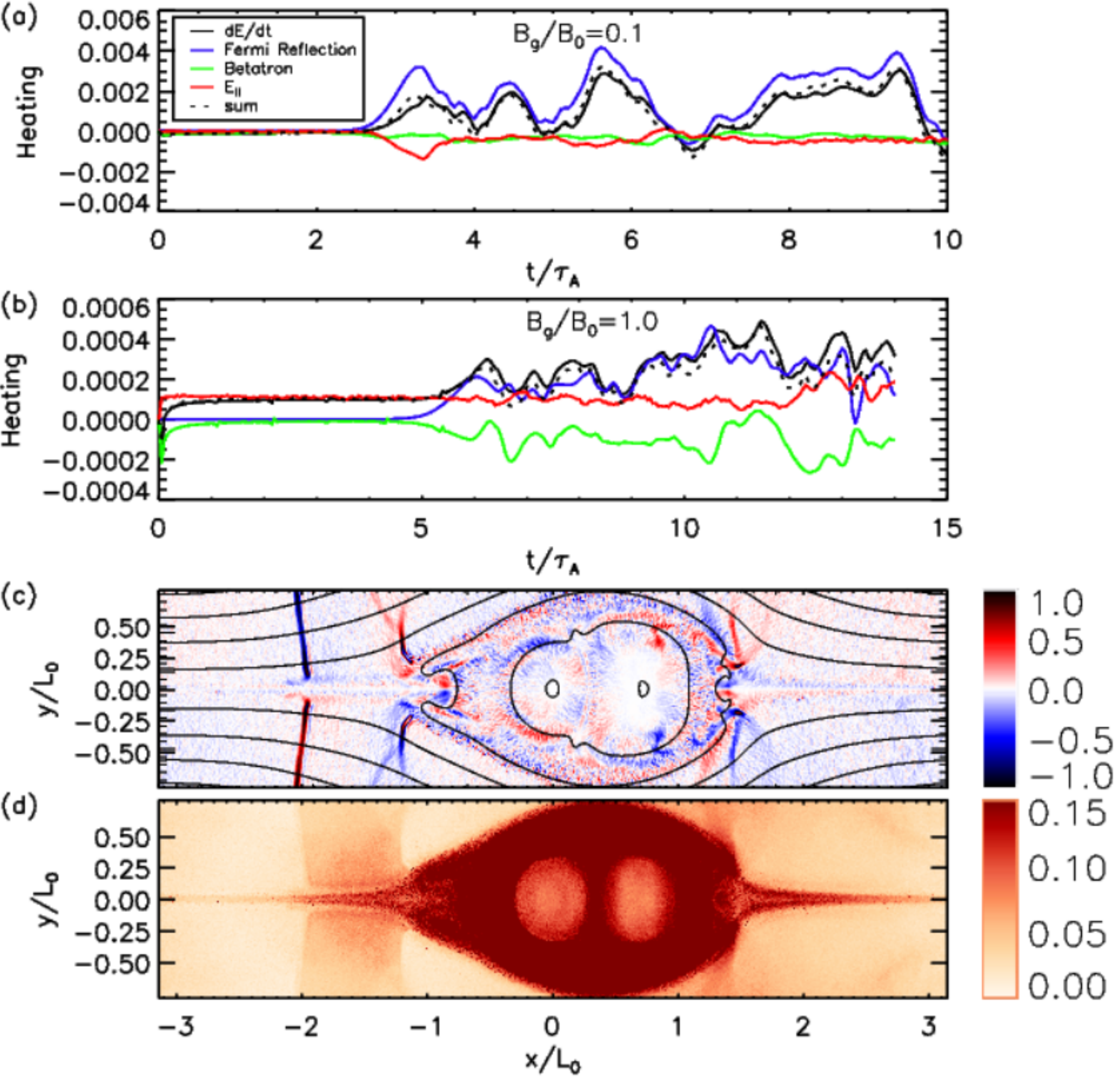}
\caption{In panel (a): the heating of the particle electrons due to Fermi reflection (blue), the large scale parallel electric field (red), betatron acceleration (green), the sum of the previous three (dashed black), and the measured heating (black) versus time for $B_g/B_0$=1.0. In panel (b): the same but for $B_g/B_0=0.1$. In panel (c) and (d): the parallel electric field with field lines overplotted and $<W_{||}>$ respectively at late time for $B_g/B_0=0.025$ and initial electron temperature $0.02$.}
\label{heating}
\end{figure}

{\bf Fitting procedure of the particle electron distributions}
To model the electron distribution functions in Fig. 4 we
use the sum of a Maxwellian and a kappa distribution. The functional
forms of the two distributions are shown in
Eq.~\ref{Supp:distfit}. Since the kappa function only has two free
parameters, we include the Maxwellian component to capture the power
law index, the thermal temperature, and the relative number of
nonthermal electrons.
\begin{equation}
    \begin{split}
    F_{fit}(W)=F_\kappa(W)+F_M(W)=\\
    \Biggl[ \frac{N_\kappa}{(\pi \kappa \theta^2)^{3/2}}\frac{\Gamma(\kappa+1)}{\Gamma(\kappa-1/2)} \left(1 + \frac{W}{\kappa \theta^2}\right)^{-(\kappa+1)} + \\
    N_M \left(\frac{m_e}{2 \pi T_M}\right)^{3/2} e^{-W/T_M} \Biggr] 4\pi \sqrt{\frac{2W}{m_e^3}}
    \end{split}
    \label{Supp:distfit}
\end{equation}
where $F_{fit}(W)$ is the fit to the total electron differential density,
$N_{\kappa}$ is the density of the kappa function, $\theta$ is the
most probable speed in the kappa function, $\Gamma$ is the Gamma
function, $N_M$ is the density of the Maxwellian function, $m_e$ is
the electron mass, $T_M$ is the temperature (in energy units) of the Maxwellian, and $W$ is the energy. Note that $\int
F_{fit}(W)dW=N_\kappa+N_M$. Since the kappa function is a Maxwellian in the
limit of low energy, we can further break up $F_{fit}$ into a
Maxwellian component, $F_M(W)$, a Maxwellian component from the kappa
function, $F_{\kappa M}(W)$, and a nonthermal component from the kappa
function that includes the power-law tail, $F_{\kappa NT}(W)$. To do
this we follow the method laid out in \cite{Oka2013}. We define a
second Maxwellian temperature, $T_{\kappa M} \equiv (1/2) m_e\theta^2$,
and set $F_{\kappa M}(T_M)=F_{\kappa}(T_M)$. This gives us an
expression for the relative density and energy of the nonthermal
electrons:
\begin{equation}
    \begin{split}
        N_{NT} = N_\kappa - N_{\kappa M} = \\
        N_\kappa \left(1-2.718 \frac{\Gamma(\kappa+1)}{\Gamma(\kappa-1/2)}\kappa^{-3/2} \left(1 + \frac{1}{\kappa} \right)^{-(\kappa+1)} \right)
    \end{split}
    \label{Supp:densities}
\end{equation}
\begin{equation}
    \begin{split}
        <W_{NT}> = <W_\kappa> - <W_{\kappa M}> = \frac{3}{2} T_\kappa  - \frac{3}{2}T_{\kappa M}
    \end{split}
\end{equation}
where $T_\kappa = (1/2)m_e \theta^2[\kappa/(\kappa-3/2)]$. $F_{fit}$
is overlaid in red on top of the particle data (in black) in the
log-log plot in Fig. 4(a). In (b) $F_{fit}$ is again
overlaid in red over the particle data, but on a linear-linear scale
zoomed in to low energies to more clearly see the thermal
population. The dual Maxwellian-kappa function fits both the low and
high energy particle data very well and can therefore be used to
explore the relative numbers of hot thermal versus nonthermal
electrons and the characteristic temperature of the hot thermals.

{\bf Analytic model of electron acceleration in a current sheet}
We present a model for electron acceleration in a current layer
with merging magnetic flux ropes that captures the essential results
of the {\it kglobal} simulations, including an expression for the
power-law index of the nonthermal electrons and its dependence on the
ambient guide field. The model includes both diffusion along the
current layer as well as the convective loss of electrons injected
into large, inactive flux ropes.

We first calculate electron energy gain during the merging of two flux
ropes of radius $r_i$, azimuthal magnetic $B_{\theta i}$ and guide
field $B_g$ as shown in Fig. 3(c). Consistent with extensive
PIC simulation results that the dominant electron heating is parallel
to the local magnetic field \cite{Dahlin2014,Dahlin2017}, we neglect
plasma compression and associated betatron acceleration. Parallel
heating results from the invariance of the parallel action $\oint
v_\parallel dl$ as merging field lines contract from the figure-eight
on the left of Fig. 3(c) to the circle on the right. Thus,
the change in the energy during the merger of two flux ropes can be
calculated by evaluating the geometry of the magnetic field before and
after the merger. For an incompressible merger, the radius of the
final flux rope is $r_f=\sqrt{2}r_i$ and the flux is preserved
$r_fB_{\theta f}=r_iB_{\theta i}$ \cite{Drake2013}. The effective
field line length of the initial state is twice the length of a single
flux rope, $s_i=2\pi r_iB_i/B_{\theta i}$, where
$B_i^2=B_g^2+B_{\theta i}^2$, since a reconnecting field line wraps
around both flux ropes as can be seen by the recently reconnected
field line in Fig. 3(c),. The final field line length is
$s_f=2\pi r_fB_f/B_{\theta f}$. Thus, invoking the invariance of the
parallel action, the final electron parallel energy $W_f$ is given by
$W_is_i^2/s_f^2$. The energy change can be re-written as a rate
equation for electron energy gain
\begin{equation}
  \overset{\boldsymbol{.}}{W}=\frac{d}{dt}W=W\frac{g}{\tau_r}, 
  \label{energydot}
\end{equation}
with $\tau_r\sim r_i/Rc_{Ax}$, where $R\sim 0.1$ is the normalized rate of merger of the flux ropes in the current layer, $c_{Ax}$ the Alfv\'en speed
based on the reconnecting magnetic field $B_x$ of the current layer, and the factor
$g=(1+2B_g^2/B_x^2)^{-1}$. The factor $g$ describes the increase
of the effective radius of curvature of the magnetic field in the
presence of a guide field, which reduces the strength of Fermi reflection
and associated energy gain during reconnection
\cite{Drake2006b,Dahlin2016,Dahlin2017}.

With the energy gain in Eqn.~(\ref{energydot}), we can write down an
equation for the number density $F(x, W, t)$ of electrons
per unit energy undergoing reconnection driven acceleration in a
one-dimensional current layer and experiencing convective loss,
\begin{equation}
  \frac{\partial}{\partial t}F+\frac{\partial}{\partial x}v_x(x)F+\frac{\partial }{\partial W}\overset{\boldsymbol{.}}{W}F-D\frac{\partial^2}{\partial x^2}F=\frac{1}{\tau_{up}}F_{up}
  \label{fdot1}
  \end{equation}
where $v_x(x)$ describes the convective loss of electrons as they are
ejected at the Alfv\'en speed out of the current layer and we include
a simple constant diffusion of electrons within the current layer. The
electrons are injected into the layer with an initial distribution
$F_{up}$ which is taken as a low-temperature Maxwellian. Although the
simulations carried out here are periodic and particles are therefore
not lost, the large flux ropes that emerge at late time and no longer
participate in the reconnection process act as sinks for energetic
electrons \cite{Dahlin2015, Li2019}.

Further, as magnetic flux continues to be added to
these flux ropes, electrons trapped in the islands become disconnected
from the current layers. As in the classical problem of diffusive
shock acceleration, the boundary condition on $f$ at the injection
point into the flux rope is zero slope.

The steady state solution of Eq.~(\ref{fdot1}) can be written as a sum
of harmonics of $F\sim\sum_n F_n\cos (n\pi x/L)$, where we take the
current layer to be centered at $x=0$ and the injection in the large
flux ropes to take place at $x=\pm L$ at the Alfv\'en speed
$c_{Ax}$. However, the problem is simplified if the diffusion $D$ is
large so that the harmonics $F_n$ with $n\neq 0$ are small. In this
limit Eqn.~(\ref{fdot1}) can simply be integrated over $x$ to obtain an
expression for $F_0$,
\begin{equation}
  L\frac{\partial}{\partial W}\frac{W g}{\tau_w}F_0+c_{Ax}F_0=\frac{L}{\tau_{up}}F_{up}.
\end{equation}
This equation is readily inverted for $F_0$,
\begin{equation}
  F_0=\frac{1}{\tau_{up}}\frac{\tau_r}{gW}W^{-c_{Ax}\tau_r/gL}\int_0^W dWW^{c_{Ax}\tau_r/gL}F_{up}(W).
  \label{f0}
\end{equation}
For low upstream temperature the energy integral can be extended to infinity and $F_0$ has the scaling
\begin{equation}
  F_0\propto \frac{1}{W}W^{-c_{Ax}\tau_r/gL} \sim W^{-(1+r_i/gRL)}.
  \label{f0scaling}
\end{equation}
The energetic spectrum takes the form of a power-law with a spectral
index that depends on the rate of reconnection $R$, the relative size of the merging flux ropes that drive electron energy gain compared with the half-width of the current sheet $L$ and the strength of the guide field.



{}

\end{document}